# Studies on proximity effect in Mo/$Bi_{1.95}Sb_{0.05}Se_3$ hybrid structure

**E. P. Amaladass[1a)], Shilpam Sharma[1], T. R. Devidas[1], Awadhesh Mani[1b)]**

[1]*Condensed Matter Physics Division, Materials Science Group, Indira Gandhi Centre for Atomic Research, Kalpakkam-603102, India*
*a) Email: edward@igcar.gov.in*
*b) Email: mani@igcar.gov.in*

Proximity effect in a mechanically exfoliated $Bi_{1.95}Sb_{0.05}Se_3$ topological insulator (TI) single crystal partially covered with disordered superconducting (SC) Mo thin film is reported. Magnetotransport measurement was performed simultaneously across three different regions of the sample viz. SC, TI and SC/TI junction. Resistance measured across SC shows a $T_C$ at 4.3 K concomitantly the resistance measurement on TI showed a metallic trend with a steep upturn at TC. Magneto-resistance (MR) measurement on TI exhibit a positive MR with Shubnikov-de Haas (SdH) oscillations, whereas on SC a positive MR superimposed with steep cusp close to $T_C$ is observed. Across SC/TI junction both SdH oscillation and the cusp were observed. The frequency of SdH oscillation on SC/TI junction is found to be lesser (~ 125 T) as compared to a reference $Bi_{1.95}Sb_{0.05}Se_3$sample (~ 174 T). Upper critical field $H_{C2}$ deduced from WHH fit was found to be 17.14 T for a reference Mo film whereas Mo film deposited on TI showed a decreased $H_{C2}$ of 4.05 T. The coherence length for the former was found to be 4.38 nm and for the latter 9.01 nm. The interaction between the spin-less Cooper pairs in SC with the spin-momentum locked carriers on the surface of TI is believed to cause such changes in transport properties.

**Keywords: Proximity effect, topological insulator, superconductor, magneto-transport, upper critical field**
**PACS: 74.45.+c, 03.65.Vf, 74.25.Op, 75.47.Np**

## I. Introduction

Bismuth based materials such as Bismuth Selenide ($Bi_2Se_3$), Bismuth Antimonide ($Bi_{1-x}Sb_x$), Bismuth Telluride ($Bi_2Te_3$) and their derivatives has been predicted theoretically and observed experimentally to be three dimensional (3D) topological insulators (TI)[1-7]. Due to strong spin-orbit coupling and topological nature of bands, these materials host two dimensional (2D) metallic surface states with Dirac-like band dispersion co-existing with an insulating bulk states. The spins of massless Dirac Fermions in these surface states are tightly coupled to their momentum and are protected by time reversal symmetry (TRS). This feature of TI make them immune to localization and backscattering, and thereby are proposed to find its way in application like spin-dependent and dissipation-less transport in spintronics devices [8,9]. Therefore, the key quantum mechanical (QM) feature of a TI, which is interesting for both basic research and technological application, emerges from symmetry protection. This is in contrast to other QM states such as superconductivity (SC) and ferromagnetism (FM) which emerge from symmetry breaking viz. gauge symmetry in former and spin symmetry in the later. The interplay between the symmetry protected TI states and symmetry broken SC and FM states thus provides rich avenues to explore new phenomena. Such investigations are amenable in FM/TI and SC/TI heterostructures; where proximity of FM or SC are believed to break the TRS in TI at the interface and give rise to novel quantum mechanical phenomenon such as anomalous quantum hall effect (AQHE), realization of Majorana Fermions and fault tolerant quantum computing [10-12].

Previous reports on TI/SC junctions showed a sharp rise in the resistance at $T_C$ when superconducting In, Al and W electrodes were used on the $Bi_2Se_3$ thin films[13] or in Nb-$Bi_2Te_3$ hybrid structure [14]. They also showed that the proximity effect of TI weakens the $T_C$ and the critical current of SC. On contrary, transport measurement on $Bi_2Se_3$ nano-ribbon and flakes using SC electrodes showed a downturn in resistance at $T_C$ and at zero magnetic field[15-17]. In order to shed more light on the interaction and proximity effect in TI/SC heterostructures, we study the transport properties of disordered Mo thin film sputter deposited on a mechanically exfoliated $Bi_{1.95}Sb_{0.05}Se_3$ TI single crystal.

## II. Experimental details

The single crystal of $Bi_{1.95}Sb_{0.05}Se_3$was grown by using high purity (99.999%) powders of Bi, Se and Sb as starting materials. Stoichiometric mixtures of elements are melted in an evacuated quartz tube at 850ºC for 24 hours followed by slow cooling at the rate of 2ºC/hour until 550ºC. After keeping at 550ºC temperature for 24 hours the sample was cooled to room temperature.The single crystals thus obtained could be easily cleaved along the basal plane leaving a silvery shiny mirror like surface. Powder x-ray diffraction was carried out on powdered single crystals at INDUS II synchrotron facility at RRCAT, Indore. Mechanically exfoliated sample was transferred on a glass substrate.

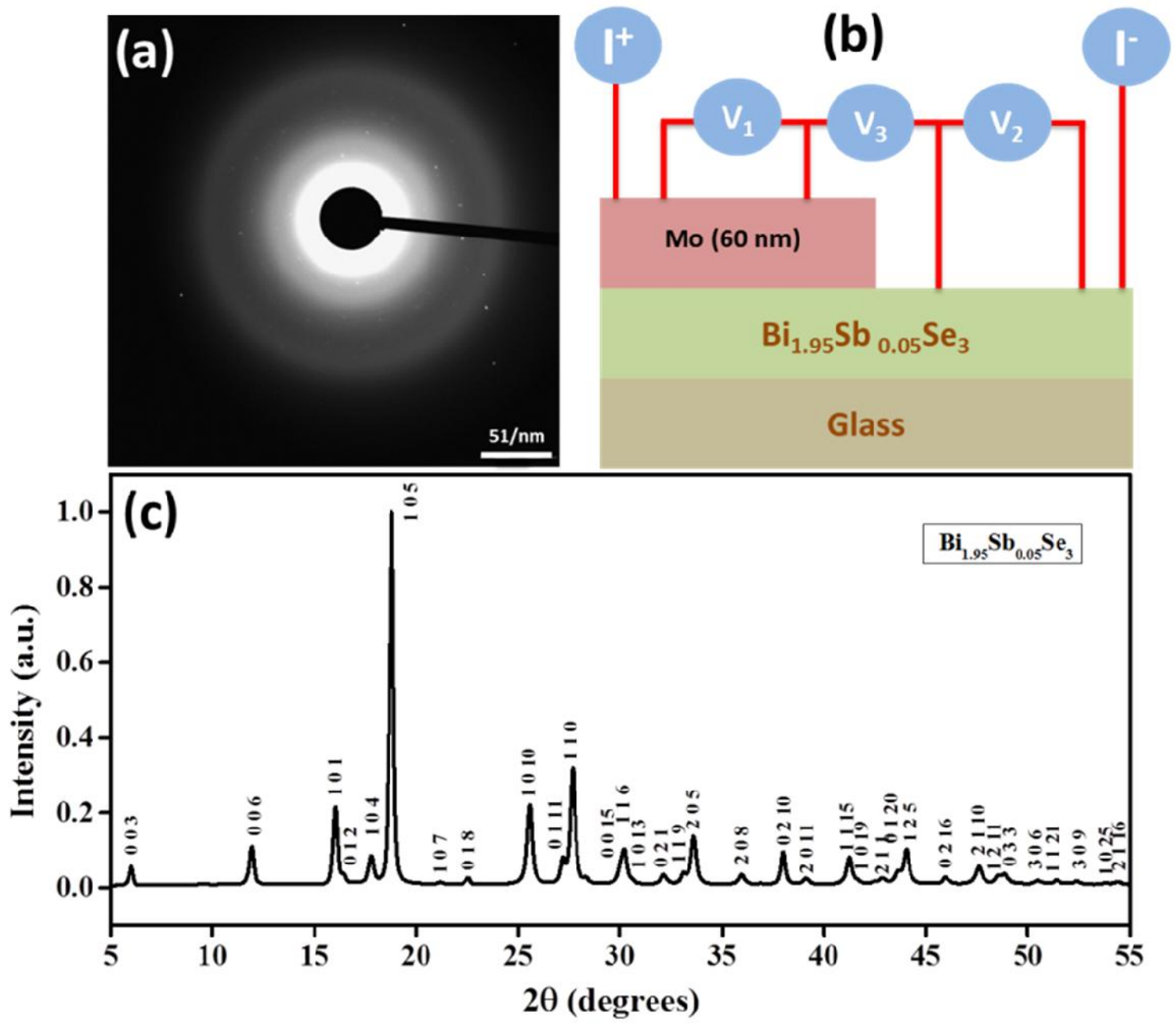

**Fig. 1.** (a) SAED pattern recorded in HR-TEM on a reference Mo film show diffused amorphous ring. (b) Shows the schematic diagram of the device structure, where current is passed across the device and the resistance is measured at three different channel $V_1$,$V_2$,$V_3$ across SC, TI and SC/TI junction, respectively. (c) Shows XRD pattern of $Bi_{1.95}Sb_{0.05}Se_3$ single crystal with all peaks indexed to rhombohedral structure and no Se/Sb/Bi impurity peaks were observed in the patterns.

A part of TI sample was shadow-masked and transferred into home built 1” magnetron sputtering system with oxygen partial pressure of < $10^{-9}$ mbar and a base pressure of ~5 x $10^{-7}$ mbar. Disordered Mo film of 56 nm thickness was deposited at 40 mA and 300 V. Argon pressure during deposition was maintained at 4.5 x $10^{-3}$ mbar and the sample to target distance was kept at 80 mm. Resistivity measurements on a reference Mo sample showed superconducting transition temperature ($T_C$) at ~ 3.7 K due its amorphous and disordered nature. The selected area diffraction pattern (SAED) in Fig. 1(a) shows diffused rings confirming the amorphous natures of the deposited films. Detailed studies on the Mo thin films with different degree of disorders are reported elsewhere[18]. The sample was then pasted on a puck holder using double sided tape. Electrical contacts were made using a 20μ Au wires and with room temperature-cured silver paste. Two current leads across the device and three voltage leads on SC, TI and SC/TI junction as shown schematically in Fig.1(b) were connected. Hereafter the resistance measured on SC region is designated as channel $V_1$, on TI as channel $V_2$ and across SC/TI junction as channel $V_3$. Temperature and magnetic field dependent resistance measurement were carried out in standard linear geometry using a commercial, 15 T Cryogen-free system from Cryogenic. Ltd, UK. With a special switching and scanning unit from Keithley Instruments, resistance at three channels $V_1$, $V_2$ and $V_3$ were measured simultaneously.

### III. Results and discussion

Temperature dependence of resistance R(T) was measured while ramping temperature from 2 K to 300 K at the rate of 2K/min. Fig.2 shows the R(T) behaviour at channel $V_1$, $V_2$ and $V_3$ respectively. R(T) on Mo thin film (channel $V_1$) shows a metallic trend and undergoes a superconducting transition at $T_C$ ~ 4.3 K. Though the R(T) across Mo/TI junction (channel $V_3$) shows a $T_C$ at same temperature ($T_C$ = 4.3 K), it does not exhibit zero resistance state. This indicates that across the junction, the $T_C$ is biased with ~30 mΩ resistance stemming from the TI and SC/TI interface (see Fig. 2). Resistance measurement on the TI (channel $V_2$) also shows a metallic behaviour from 300 K down to 6 K. Interestingly, a resistance minimum is observed at 4.3 K, which exactly corresponds to the $T_C$ of Mo thin film (see the inset of Fig. 2). Below 4.3 K, resistance increases gradually with decreasing temperature as shown in the inset of Fig. 2. This anomalous behaviour is reported [19] to arise due to the incompatible nature of the carriers in superconductor with the surface carriers in TI. In an SC, the Cooper pair encompasses electron with antiparallel spins whereas on the surface of a TI the spin of the electron is locked to its momentum and therefore they align parallel to each other. The incompatibility of these quantum mechanically distinct carriers gives rise to the resistance enhancement at $T_C$. However, the order of resistance change in our sample at $T_C$ is very less (0.2 mΩ) as compared to previous reports[13,14] which might be due to different geometry of measurements employed. Due to the fact that resistance upturn is masked by the superconducting transition as seen in channel $V_1$ and $V_3$, the upturn in resistance is observed only in channel $V_2$. Resistance at channels $V_1$, $V_2$ and $V_3$ as a function of magnetic field in the range of -15T<0T<15T and in temperature range of 2 -7 K is shown in Fig. 3(a)-(c), respectively. At 7 K, above the $T_C$ of Mo films, MR from channel $V_1$ shows a positive behaviour. At 2 K it shows a drastic decrease in resistance between -4 T and +4 T and attains a zero resistance state at 0 T field. On further increasing the temperature to 3 K, MR shows similar positive dip but at reduced field range between -2.5 T and 2.5 T. A shallow dip without zero state resistance appears at 4.2 K between -0.5 T and 0.5T.

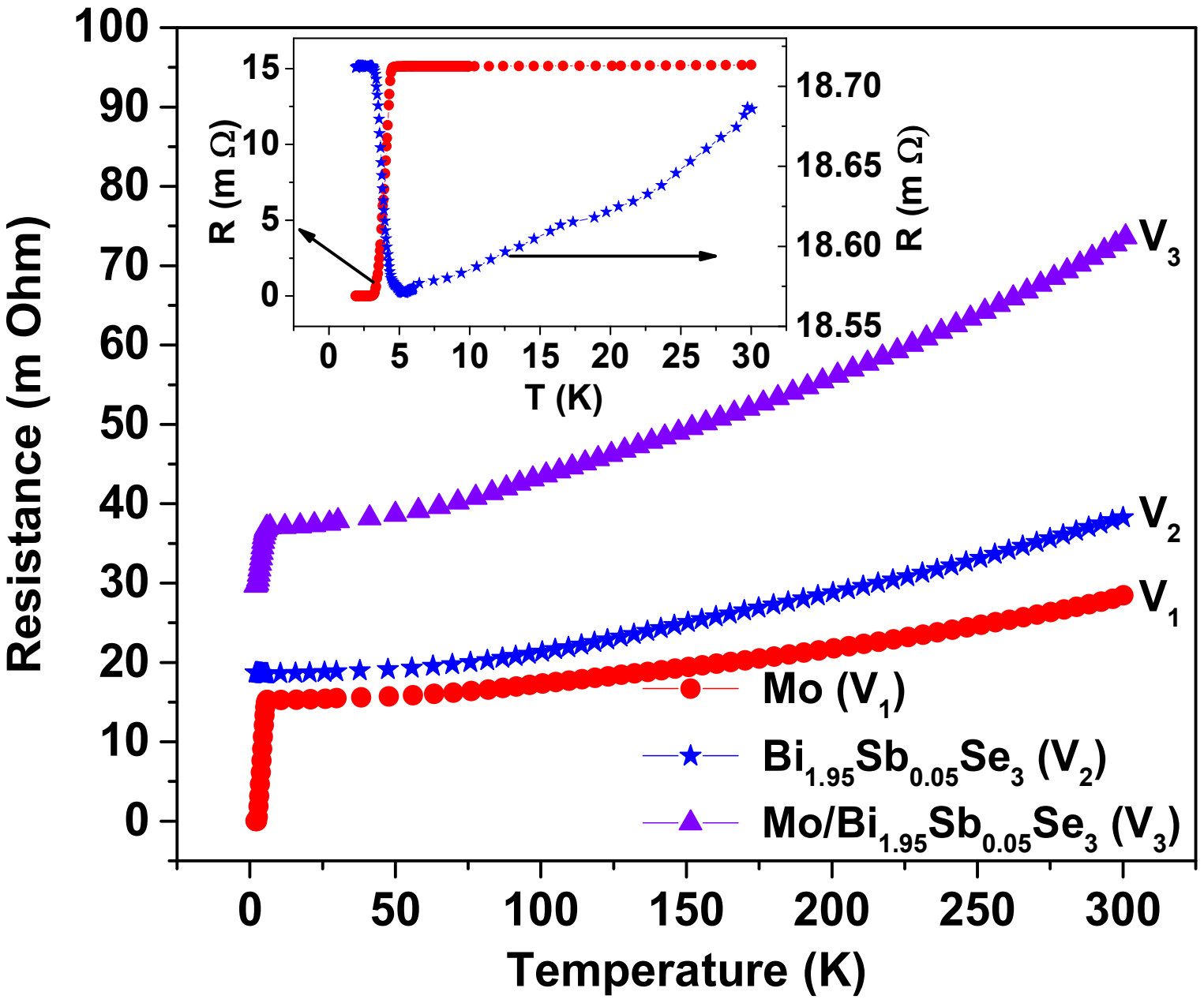

**Fig. 2.** Temperature dependence of resistance across Mo (channel $V_1$), Topological insulator $Bi_{1.95}Sb_{0.05}Se_3$ (channel $V_2$) and across SC/TI junction (channel $V_3$). Inset shows the upturn in the resistance measured on TI (channel $V_2$) at the $T_C$ of Mo superconductor (channel $V_1$)

This MR dip is attributed to the temperature and field dependent critical field behaviour of the SC. The resistance measured in channel $V_2$ in presence of magnetic field was noisy as compared to the temperature dependent R(T) behaviour at zero field in the same channel. Shubnikov-de Haas (SdH) oscillations are observed at higher field, $B \geq \pm 5$ T. At 2 K and 3 K, it showed a linear positive MR slope between ± 0.5 T in contrast to positive MR dip seen in other channels. The origin of this anomalous behaviour seen in channel $V_2$ is not clear. MR measured across the SC/TI junction (channel $V_3$) shows similar positive dip as seen in channel $V_1$. In addition, it exhibits SdH oscillation for fields $B \geq \pm 5$ T similar to $V_2$. The frequency of oscillation was deduced using Fast Fourier Transform (FFT) analysis of the resistance versus 1/B plot. Fig. 3(d)-(f) shows the frequency of SdH oscillation in a reference TI sample to be 174 T whereas in channel $V_2$ and $V_3$ are 122 T and 128 T, respectively. A detailed study on $Bi_{2-x}Sb_xSe_3$ sample for different Sb content showed the SdH oscillation from $Bi_{1.95}Sb_{0.05}Se_3$ are from surface states which has been inferred from analysis in terms of Lifshitz-Kosevich equation and Landau level fan diagram[20]. Therefore SdH oscillation emerging from the two dimensional electron gas (2DEG) at the bulk/surface interface [21] is ruled out. Due to the fact that the frequency F is related to the area ($A = \pi k_F^2$) of 2D Fermi surface via Onsager relation $F = (\hbar/2\pi e)A$, the decrease in the frequency in channel $V_2$ and $V_3$ is attributed to downward shifting of Fermi level towards the Dirac point. This is surprising because if we presume that the surface of the TI would have deteriorated during device fabrication, the Fermi level should have moved away from Dirac point[22]. Hence, it lends us to believe that the difference in the SdH oscillation frequencies is solely due to the proximity of SC which has altered the surface characteristics of the TI.

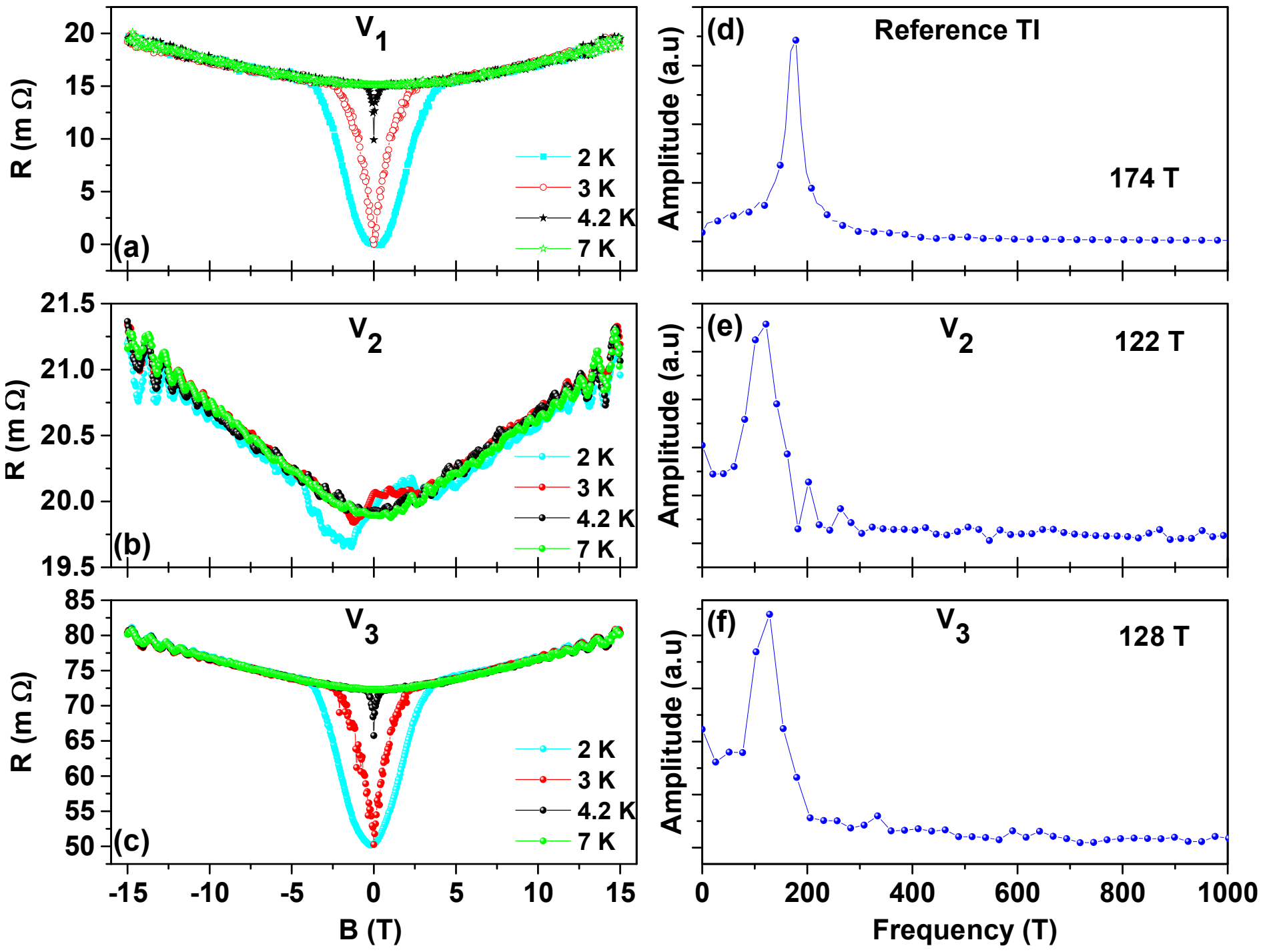

**Fig. 3.** Resistance as function of perpendicular magnetic field (a) channels $V_1$,(b) channels $V_2$ and (c) channels $V_3$. FFT analysis of the SdH oscillation in a reference $Bi_{1.95}Sb_{0.05}Se_3$ sample (d), channel $V_2$ (e) and channel $V_3$ (f).

In order to further investigate the effect of interplay of TI ($Bi_{1.95}Sb_{0.05}Se_3$) properties on the critical field behaviour of the SC, we carried out R(T) measurement in 2 K -300 K temperature range at different magnetic fields ranging from 0 T to 15 T. A criterion that the temperature at which value of resistance falls to 90% of its normal state prior to superconducting transition was used to estimate the $T_C$ at each applied field. Zero-field superconducting transition temperature of reference Mo sample was found to be ~3.7K. The plots of resistance in channel $V_1$ and $V_3$ as a function of temperature measured under magnetic fields up to 15 T, applied perpendicular to the surface of heterostructure is shown in Fig 4. A positive magneto-resistance has been observed in the normal state of disordered Mo film deposited on to TI single crystal. It should be mentioned that there was no appreciable magneto-resistance in the normal state of the reference Mo sample deposited simultaneously on a glass substrate (not shown). This indicates that the proximity of TI affects even the normal state electronic properties of Mo thin film. The superconducting transitions temperature decreases with increasing the magnetic field in channel $V_1$ and $V_3$ as shown in the inset of Fig. 4.

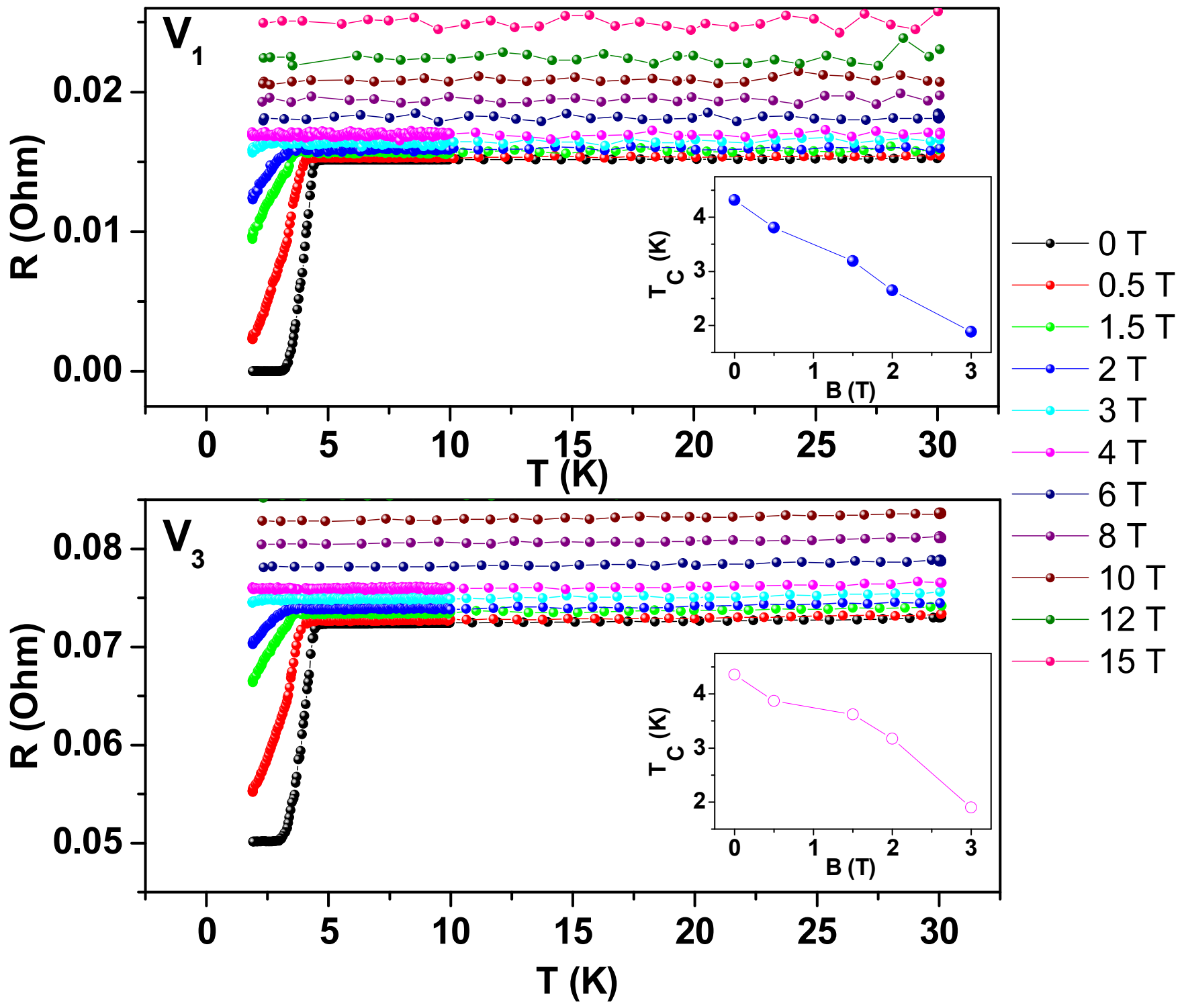

**Fig. 4.** Resistance as a function of temperature measured under different magnetic fields up to 15 T for channel $V_1$ and channel $V_3$. Insets show the variation in $T_C$ with applied field.

The temperature dependence of upper critical fields ($H_{C2}$) in the plane of Mo thin films deposited on TI as well as on the glass is shown in Fig. 5. From the Fig it is clear that the Mo deposited on the glass shows typical type II behaviour with its upper critical field varying as BCS superconductor. This cross-over of Mo from a usual type I behaviour to type II behaviour could be understood as disorder induced increase in penetration depth ($\lambda$) and decrease in coherence length ($\xi$)[23]. The variation of upper critical as a function of reduced temperature has been fitted to $H_0(1\text{-}t)^n$, where $t = T/T_C$. The value of $H_0$ for reference sample turn out to be ~ 16.4 T and n ~ 0.72. This is much higher than the critical field of bulk Mo ~ 98 mT. The temperature variation of critical field for Mo deposited on TI shows a linear behaviour. A Similar fit to $H_0(1\text{-}t)^n$ yielded $H_0$ ~ 5.4T and n ~1.02. The critical field $H_0$ for Mo on TI is lower than the critical field estimated for the reference Mo sample. This could be due to effect of spin locked states existing on the surface of TI. The $H_{C2}(T)$ data was fitted to WHH formalism[24].

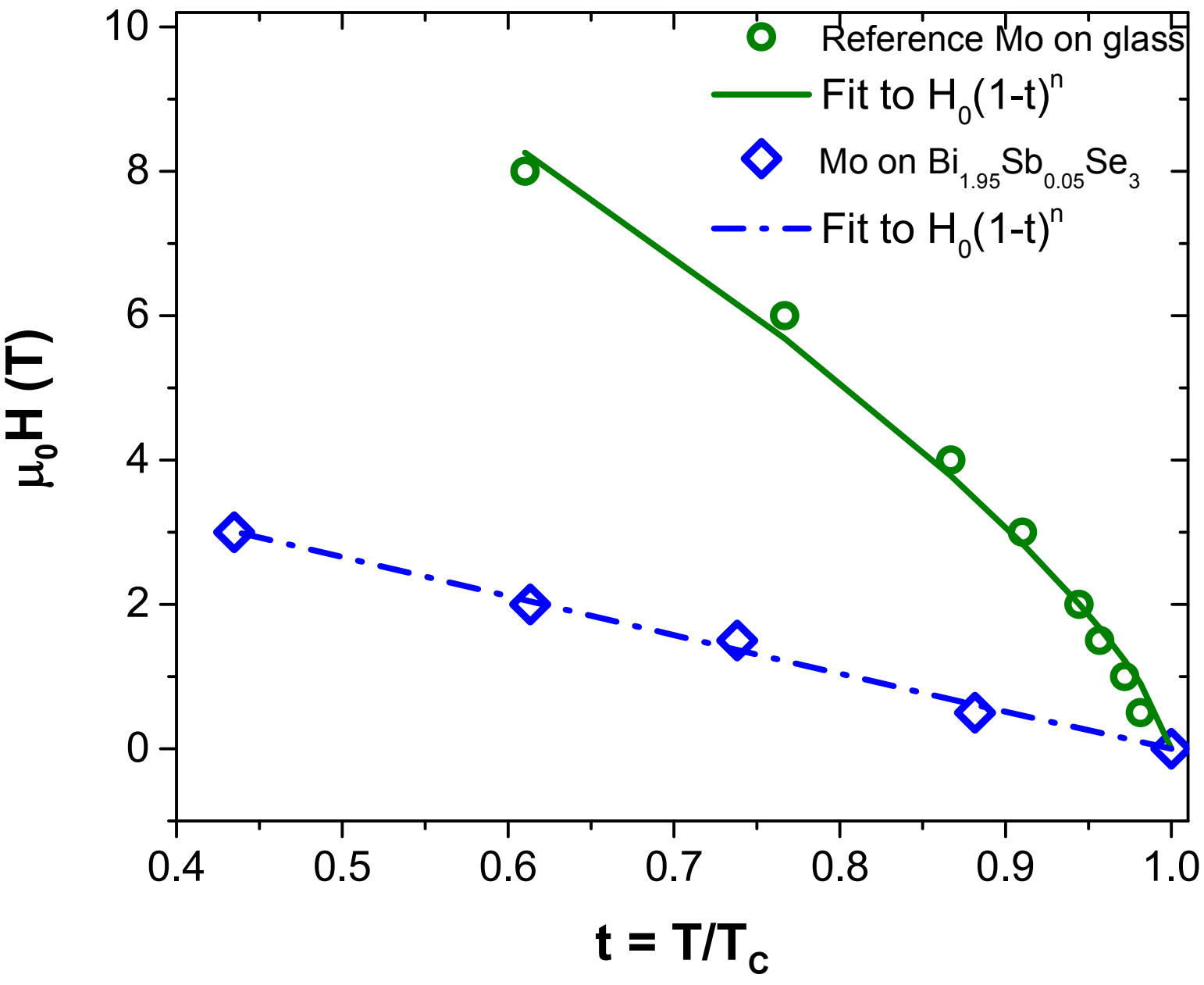

**Fig. 5.** Comparison of the temperature dependence of upper critical fields ($H_{C2}$) of Mo thin films deposited on TI to a reference Mo thin film deposited on a glass substrate with its fits.

The slope $\frac{dH_{C2}}{dT}|_{T_C}$, Maki parameter ($\alpha$) and spin-orbit coupling term ($\lambda_{SO}$) were used as the fitting parameters. For the reference sample satisfactory fit to WHH formula was not found whereas $H_{C2}(T)$ for heterostructure was observed to obey WHH formalism with $\alpha = 0$. The fitted value of slope was estimated to be ~6.65 T/K and ~1.36 T/K for reference Mo and Mo on TI samples respectively (see Fig.5). The dirty limit, orbital effects limited upper critical field at zero temperature has been estimated to be ~17.14 T and 4.05 T for reference and heterostructure respectively. The zero temperature coherence length $\xi(0) = \sqrt{\frac{\phi_0}{\pi H_{C2}(0)}}$ (where $\phi_0$ is flux quantum) is seen to increase from 4.38 nm for reference Mo film to 9.01 nm in heterostructure. In nutshell, the proximity of TI is seen to affect significantly the critical properties of Mo superconductor.

## IV. Summary and conclusion

In conclusion, the interplay between the superconducting and topological properties in a SC/TI heterostructure has been studied using temperature and magnetic field dependent electronic transport measurements. It is observed that the proximity of $Bi_{1.95}Sb_{0.05}Se_3$ TI significantly alters the normal as well as superconducting properties of disordered Mo superconductor, while the proximity latter brings about a perceptible change in the electronic properties of the TI ($Bi_{1.95}Sb_{0.05}Se_3$) as well. An upturn in the resistance in the TI film is observed in the vicinity of the $T_C$ of the SC film. The frequency of the 2D SdH oscillation in the SC/TI heterostructure decreases as compared to the reference TI sample. On the other hand, the $H_{C2}$decreases ($\xi$ increases) for the SC on TI as compared to the reference SC sample on glass substrate. The scattering and the interaction between the spin-singlet Cooper pairs with the spin polarised carriers in TI is believed to cause such changes in the transport properties of the SC/TI junction.

**Acknowledgement**

The author acknowledges Dr. S. Amirthapandian for carrying out SAED measurement and Dr. A. K. Sinha for providing access to the BL-12 beam-line of Indus II, RRCAT – Indore to carry out synchrotron XRD measurements. The authors gratefully acknowledge UGC-DAE-CSR Kalpakkam node for providing access to the 15 T cryogen free magneto-resistance facilities.